\documentclass[reprint,amsmath,amssymb,aps,pre]{revtex4-1}

\usepackage{graphicx}
\graphicspath{plots}
\usepackage{dcolumn}
\usepackage{bm}
\usepackage{color}
\usepackage{bbold}
\usepackage{soul}

\usepackage{empheq}
\usepackage{flafter}     

\usepackage{catchfile}
\newcommand{\getenv}[2][]{%
\CatchFileEdef{\temp}{"|kpsewhich --var-value #2"}{\endlinechar=-1}%
\if\relax\detokenize{#1}\relax\temp\else\let#1\temp\fi}

\getenv[\articlestoragepath]{ARTICLE_STORAGE}
\IfFileExists{pdf_article_link.sty}{\usepackage{pdf_article_link}}

\newcommand{\HIDDEN}[1]{}

\usepackage{color}

\newcommand{\overbar}[1]{\mkern 1.5mu\overline{\mkern-1.5mu#1\mkern-1.5mu}\mkern 1.5mu}

\let\OLDthebibliography\thebibliography
\renewcommand\thebibliography[1]{
  \OLDthebibliography{#1}
  \setlength{\parskip}{0pt}
  \setlength{\itemsep}{0pt plus 0.3ex}
}

\makeatletter
\let\Hy@backout\@gobble
\makeatother

\begin{document}

\title{Controlling synchronization in coupled area-preserving maps using 
stickiness}

\author{Swetamber Das}
\affiliation{Max-Planck-Institut f\"ur Physik komplexer Systeme, N\"othnitzer
    Stra\ss{}e 38, 01187 Dresden, Germany}
\email{swetdas@pks.mpg.de}

\date{\today}

\begin{abstract}
Unidirectionally coupled area-preserving maps with a mixed phase space may 
show identical synchronization in the sticky neighborhood of the regular 
islands. We use this fact to devise numerical procedures to 
control (delay and expedite) the process of synchronization in two standard maps coupled under  the Pecora-Carroll coupling scheme. The delay method is based on controlled kicking of trajectories away from synchronization traps for as long as necessary. The method to expedite the process is achieved by a parameter perturbation technique which rapidly drives the chaotic trajectories to synchronization traps in the sticky neighborhoods 
of regular islands. We also discuss the limitations of these methods.
\end{abstract}

\pacs{PACS here}

\maketitle

\section{Introduction} 
A low-dimensional Hamiltonian system commonly exhibits a mixed phase space 
i.e. regular structures and chaotic regions may co-exist at a given degree of 
nonlinearity. This mixed nature has interesting consequences for the transport
properties of such systems, for instance, the existence of anomalous kinetics,
L\'{e}vy processes and L\'{e}vy flights \cite{Klafter1994,Zaburdaev2015}, 
power law contributions to recurrence and other statistics, and the existence 
of dynamical traps \cite{Zaslavsky2002a,Zaslavsky2002b}. An intriguing region 
of a mixed phase space is the interface between regular regions and chaotic 
sea \cite{Mackay1984, Easton1993}. The dynamics in these neighborhoods are 
complex but fairly well understood \cite{Meiss2015} for low dimensional 
systems. The complexity arises from the fact that a chaotic trajectory spends 
an arbitrary long but finite time at the boundaries of regular islands before 
exiting to the chaotic sea. The intermittent tendency of chaotic trajectories 
to stay close to the regular boundaries is called stickiness. A major 
consequence of the stickiness is  the existence of power law in the 
Poincar{\'e} recurrence times indicating algebraic decay for long 
times rather than exponential decay expected for normal transport. 
Therefore, due to stickiness, even a small regular island can influence the 
global transport properties of the system and decay of correlations. The 
phenomenon has been of great interest and continues to be studied on 
theoretical level \cite{Altmann2006, Altmann2010,Livorati2012,Bunimovich2012, 
Kruger2015, Lange2016}. In addition, stickiness has found application in 
several problems such as particle advection in fluids 
\cite{Babiano1994,Tel2005}, transport in 
plasma fusion devices \cite{Szezech2012,Martins2014}, celestial mechanics 
\cite{Efthymiopoulos1999, Harsoula2010, Harsoula2016}.

A surprising feature of stickiness has emerged in our earlier 
work~\cite{Mahata2016}, we have looked at the effects of the mixed phase space 
on the synchronization in a system of two standard maps coupled in 
unidirectional drive-response configuration. We have shown that 
synchronization of chaotic trajectories of the drive and response maps 
typically occurs in the neighborhood of regular inlands as a consequence of 
stickiness in the region.  This is the first instance, as far as we are aware, 
where stickiness have been found to influence synchronization in coupled 
chaotic systems. It is important to point out that a possible role of stickiness in 
Hamiltonian systems for synchronization was already predicted by 
Zaslavsky~\cite{Zaslavsky2002b} in beginning of the last decade.  For a chaotic 
orbit, synchronization typically happens via an intermittent behavior in the 
phase difference of the drive and response maps.   The sticky neighborhoods of 
a regular islands {\it temporarily} traps chaotic orbits. In such trapping 
regions in the phase space, parts of a  chaotic trajectory are almost regular 
in time and allow for synchronization to occur. Such traps are characterized 
using the properties of the finite-time Lyapunov exponent \cite{Szezech2005}. 
In the context of our work, we will refer to these traps as synchronization 
traps. We further note that the behavior of the synchronization traps can be 
analyzed in a more quantitative way by analyzing the location and stability 
properties of the periodic orbits at the locations where 
synchronization takes place.

Synchronization in coupled chaotic systems, coupled map lattices and networks 
have been studied in details (for reviews, see 
\cite{Pecora1998, Boccaletti2002, Pecora2015}). 
Coupled Hamiltonian systems are also known to show synchronization such as 
Measure synchronization~\cite{Hampton1999,Wang2003,Vincent2005,Gupta2017} and 
identical synchronization~\cite{Mahata2016,Das2017}. However, the ability to 
control synchronization based on the dynamics of the process has not been 
given much attention, until recently~\cite{Grabow2011,Wang2016}. In realistic 
systems, the speed of synchronization may be significant. For example, in 
neurosciences,  the speed of the visual and olfactory processing are 
interesting problems \cite{Thorpe1996,Uchida2003}.

In this paper, we intend to develop a couple of mechanisms to control 
synchronization process i.e. to increase and to reduce synchronization 
times based on the location of synchronization traps in the sticky 
neighborhood of regular islands in the phase space of the area-preserving 
system of the standard map. Our delay method is based on the 
knowledge of the sticky neighborhoods of a mixed phase space. A simple kicking 
mechanism is used to keep the drive trajectory away until the control 
mechanism is active. The system synchronizes when the control is removed. On 
the other hand, the method to reduce synchronization times is based on 
parameter perturbation which is a well-known mechanism for chaos 
control~\cite{Boccaletti2000, Ott2008, Bolotin2009}. Our method may be viewed 
as an adaptation of the Ott-Grebogi-Yorke (OGY) method~\cite{Ott1990} suitable 
for our unidirectionally coupled system. The OGY method is applicable, in 
principle, to a chaotic system for which a mathematical description is 
unavailable. This procedure seeks to stabilize a desired low-order unstable 
periodic orbits embedded in a chaotic attractor via time-dependent parameter 
perturbations. In another method proposed by Lai and 
Grebogi~\cite{Lai1993a,Lai1994}, two identical systems are synchronized 
through parameter perturbation. This method introduces coupling in the system 
only when parameter is perturbed. Another approach~\cite{Nagai1996} uses an  
adaptive parameter perturbation at every iteration in order to synchronize two 
coupled system. However, our technique of perturbation is similar to the one 
proposed by Aston and Bird and by Lai {\it et. al.}~\cite{Lai1993b}. These 
methods are suitable extensions of the OGY method. In the former work, a 
parameter perturbation is used for a dissipative system to constrain chaotic 
trajectories in the vicinity to the synchronous manifold. The later work 
describes a chaos control method for a  Hamiltonian system by stabilizing 
chaotic orbits around some 
desired unstable period orbit. The basic idea is to tune the parameter 
judiciously when a chaotic trajectory visits the neighborhood of a fixed or 
periodic point embedded in a chaotic attractor so as to restrict it in the 
vicinity. We also apply the perturbation locally to one of the periodic 
points which eventually forces the drive trajectory to a synchronization 
trap. 

We demonstrate the mechanism at a specific value of the 
nonlinearly parameter (see Sec.~\ref{Master stability function}) where only 
two small period 2 islands  exist in the phase space. 
\begin{figure}[t]
	\includegraphics[scale=.45]{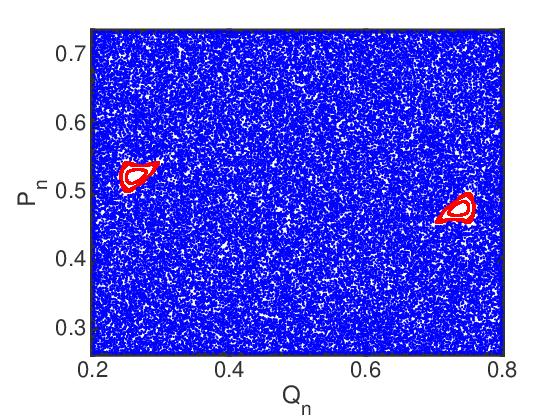}
	\caption{\label{fig:Standard_map} \footnotesize The phase space of the standard map at the parameter value K = 6.0. A couple of regular islands (shown in the color red) and a connected chaotic region (shown in the color blue). }
\end{figure}
The rest of the paper is organized in the following way: we explain the 
coupling scheme in Sec.~\ref{sec:Pecora_Carroll} and the mechanism based on 
the location of synchronization traps is given in Sec.~\ref{Master stability 
function}. The numerical procedures to delay and advance synchronization 
times are demonstrated in Sec.~\ref{sec:delay} and Sec.~\ref{sec:advanced} 
respectively, and the paper ends with conclusions in 
Sec.~\ref{sec:conclusions} including a discussion on limitations and 
implications.

\section{The Pecora Carroll's unidirectional coupling scheme}
\label{sec:Pecora_Carroll} 
The standard map is considered to be the prototypical example of a two-dimensional area-preserving map, and is given by:
\begin{empheq}[right=\empheqrbrace \mod 1 .]{align} {\label{equ:stdmap}}
P_{n+1} = P_n +  \frac{K}{2\pi}\sin(2\pi Q_n) \nonumber\\
Q_{n+1} = P_{n+1} + Q_n 
\end{empheq}

\begin{figure}[t]
	\includegraphics[scale=.3]{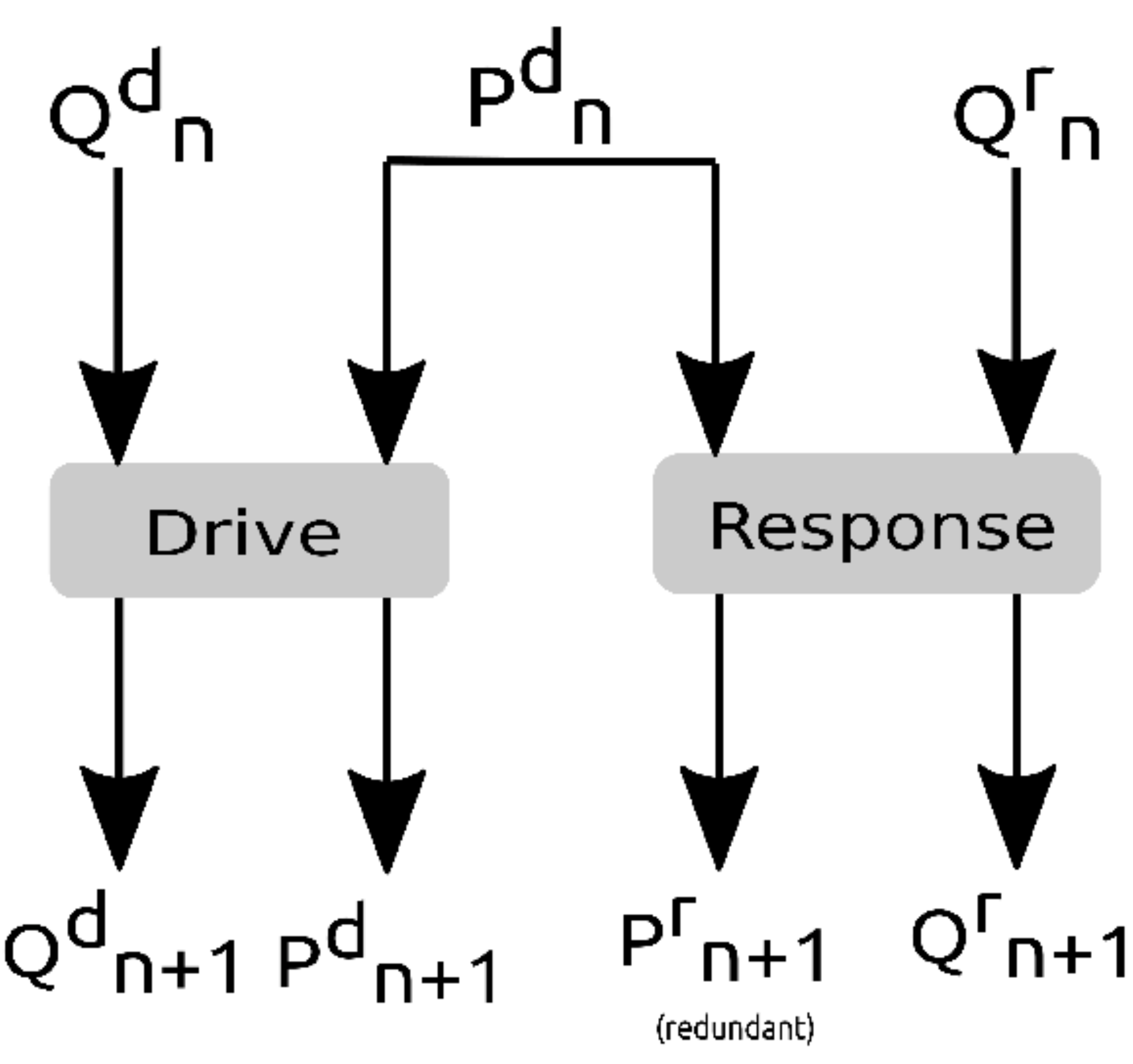}
	\caption{\label{fig:schematic} \footnotesize Schematic displaying the Pecora-Carroll method of coupling two standard map as explained in the text.}
\end{figure}
\noindent Here the subscript $n$ denotes the discrete time and $K$ is the 
nonlinearity parameter. These equations typically describe the evolution of 
two canonical variables $P$ and $Q$ which correspond to the momentum and 
co-ordinate in the Poincar\'{e} section of a freely moving rotator with 
interleaved periodic kicks. This system represents the behavior of a variety 
of systems such as charged particle confinement in  mirror magnetic traps, 
particle dynamics in accelerator, comet dynamics in solar systems 
etc.~\cite{Chirikov1960,Izraelev1980,Chirikov1989,Chirikov2008}
Two-dimensional  phase space plots  of the standard map for the parameter 
value $K = 6$ using a chaotic and a few regular initial conditions  are shown in Fig.~\ref{fig:Standard_map}.

We now synchronize two standard maps, using the Pecora-Carroll scheme of synchronization using drive-response coupling \cite{Pecora1990,Pecora2015}. 
This system was first devised to synchronize the chaotic trajectories dissipative chaotic dynamical systems. Under this unidirectional coupling scheme, we duplicate the given map and couple the original and the duplicated map in a drive-response configuration. This means that the drive map evolves freely but the evolution of the response map is dependent on the drive. In this case, the $P$ value of the response system is set to the $P$ value of the drive system, at each iterate.  A schematic is given to demonstrate the coupling scheme in Fig.~\ref{fig:schematic}.
\begin{figure}[b]
    \includegraphics[scale=.4]{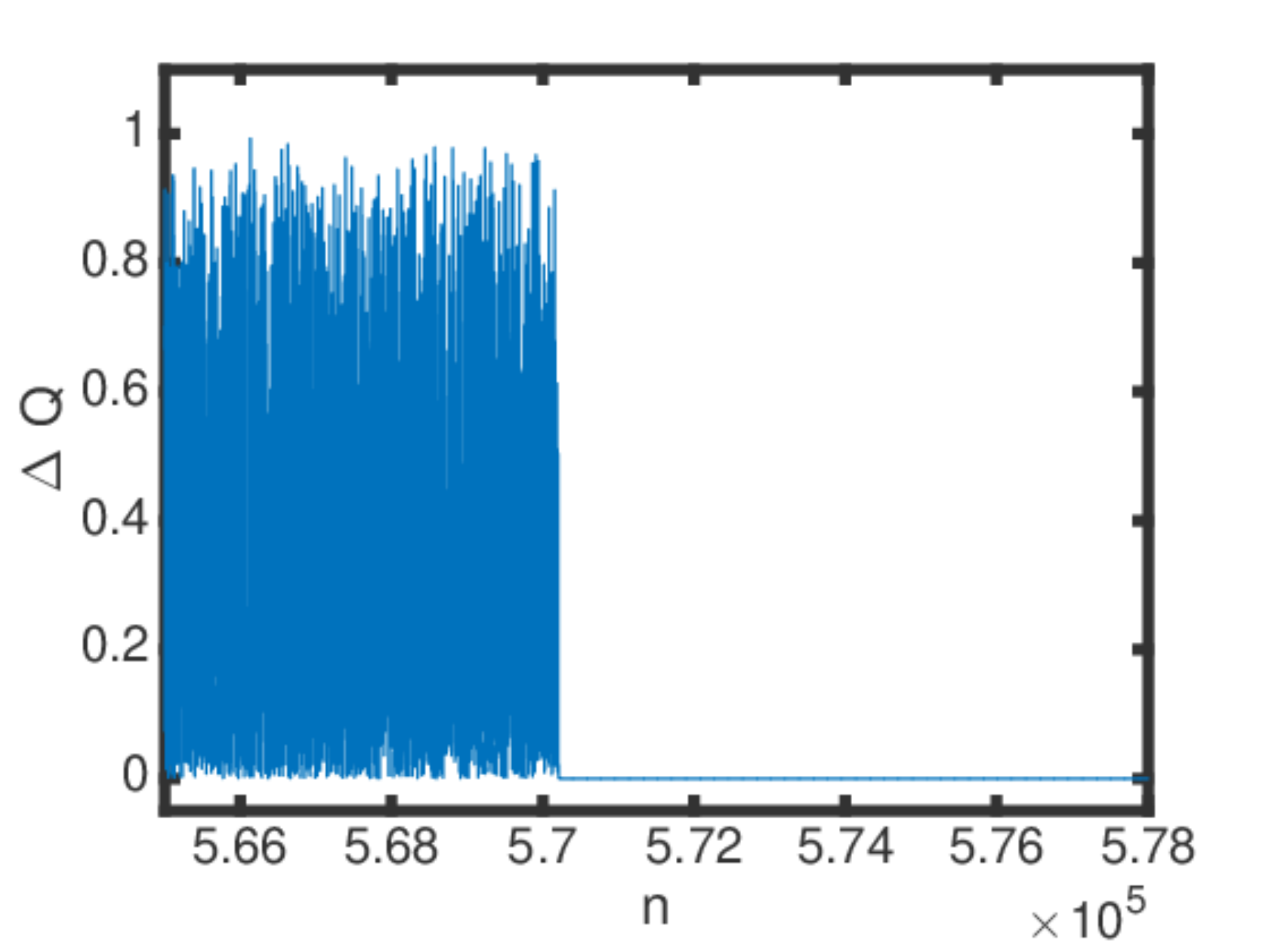}
    \caption{\label{fig:sync_ex} \footnotesize The variation of $\Delta Q = 
    Q^d_n-Q^r_n$ of the coupled system at $K = 6$ with iterations $n$. 
    Synchronization is considered to be achieved when $\Delta Q < 10^{-5}$ and 
    remain so for the rest of the iterations. The iteration step $n$ at which 
    synchronization occurs is called synchronization time $\tau$. For this 
    case, $\tau = 570 246.$ }
\end{figure}

The initial values of $Q$ in the drive and response maps are chosen 
arbitrarily, whereas the $P$ values are identical. The system is said to reach 
complete synchronization when both the $Q$ values of the drive and response 
systems evolve identically i.e. 
\begin{equation}
\lim_{t\rightarrow\infty}(Q^\textnormal{d}_n - Q^\textnormal{r}_n) = 0.
\end{equation}

Synchronization time is the least value of the iterate say, $\tau$, where $\Delta Q = Q^d-Q^r$ vanishes for all subsequent iterations: 
\begin{equation}
(Q^\textnormal{d}_n - Q^\textnormal{r}_n) = 0; \hfill  n \geq  \bf {\tau }
\end{equation}

In all of the computations reported in the work, two chaotic trajectories are considered to be synchronized to numerical accuracy, if the Euclidean distance between them is less than $10^{-5}$. An example is shown in 
Fig.~\ref{fig:sync_ex}. Here, we plot $\Delta Q = Q^\textnormal{d}_n-Q^\textnormal{r}_n$ for initial 
conditions $(P^\textnormal{d}_0,Q^\textnormal{d}_0,Q^\textnormal{r}_0) = 
(0.569, 0.906,0.106)$ at $K = 6$. The synchronization time is found to be 5,70,246.

We now analyze synchronization of the coupled system using the master stability function~\cite{Pecora1998}. Numerical procedures to control synchronization times will be discussed thereafter.  

\section{Locating Synchronization Traps using Master stability function}
\label{Master stability function}
A general drive-response system coupled unidirectionally may be described by the following set of equations:

\begin{align}
\frac{d\overbar{X}_\textnormal{d}}{dt} & = \overbar{F}(\overbar{X}_\textnormal{d}) \nonumber  \\
\frac{d\overbar{X_\textnormal{r}}}{dt} &= \overbar{F}(\overbar{X}_\textnormal{r}) + \alpha 
E(\overbar{X}_\textnormal{d}-\overbar{X}_\textnormal{r}).
\end{align}

\noindent Here $\overbar{X}_d$ and $\overbar{X}_r$ are drive and response 
variables; the matrix $E$ determines the linear combination of the 
$\overbar{X}$ used in the difference and $\alpha$ is the coupling strength. 
For the map case, we have the following form

\begin{align}
\overbar{X}^\textnormal{d}_{n+1} &= \overbar{F}(\overbar{X}^\textnormal{d}_{n}) \nonumber  \\
\overbar{X}^\textnormal{r}_{n+1} &= \overbar{F}(\overbar{X}^\textnormal{r}_{n}) + \alpha 
E(\overbar{X}^\textnormal{d}_{n}-\overbar{X}^\textnormal{r}_n).
\end{align}

Therefore, in the case of a general unidirectional coupling of two standard maps, we get
\begin{align}
P^\textnormal{d}_{n+1} &= P^\textnormal{d}_n + \frac{K}{2\pi}\sin(2\pi Q^\textnormal{d}_n) \nonumber\\
Q^\textnormal{d}_{n+1} &= P^\textnormal{d}_{n+1} + Q^\textnormal{d}_n \nonumber\\
P^\textnormal{r}_{n+1} &= P^\textnormal{r}_n + \frac{K}{2\pi}\sin (2\pi Q^\textnormal{r}_n) + \alpha(P^\textnormal{d}_n-P^\textnormal{r}_n) \nonumber \\
Q^r_{n+1} &= P^r_{n+1} + Q^r_n.
\label{equ:coupled}
\end{align}

We have chosen $E$ to be the matrix $\begin{bmatrix} 1 & 0 \\ 1 & 0 \end{bmatrix}$. 

To find the stability of the synchronous state, we first express  
Eq.(\ref{equ:coupled}) in terms of  $P^\perp = P^\textnormal{d} - 
P^\textnormal{r}$ and $Q^\perp = Q^\textnormal{d} -Q^\textnormal{r}$, as 
follows
\begin{eqnarray}
\label{equ:trans}
P^\perp_{n+1} &= (1-\alpha)P^\perp_n+ \frac{K}{2\pi} \sin(2\pi Q^\textnormal{d}_n) - 
\frac{K}{2\pi} \sin(2\pi Q^\textnormal{r}_n) \nonumber \\ 
Q^\perp_{n+1} &= (1-\alpha)P^\perp_n+ Q^\perp_n + \frac{K}{2\pi} \sin(2\pi 
Q^\textnormal{d}_n) \nonumber \\ &- \frac{K}{2\pi} \sin(2\pi Q^\textnormal{r}_n).
\end{eqnarray}

\begin{figure*}[t!]
	\includegraphics[scale=0.41]{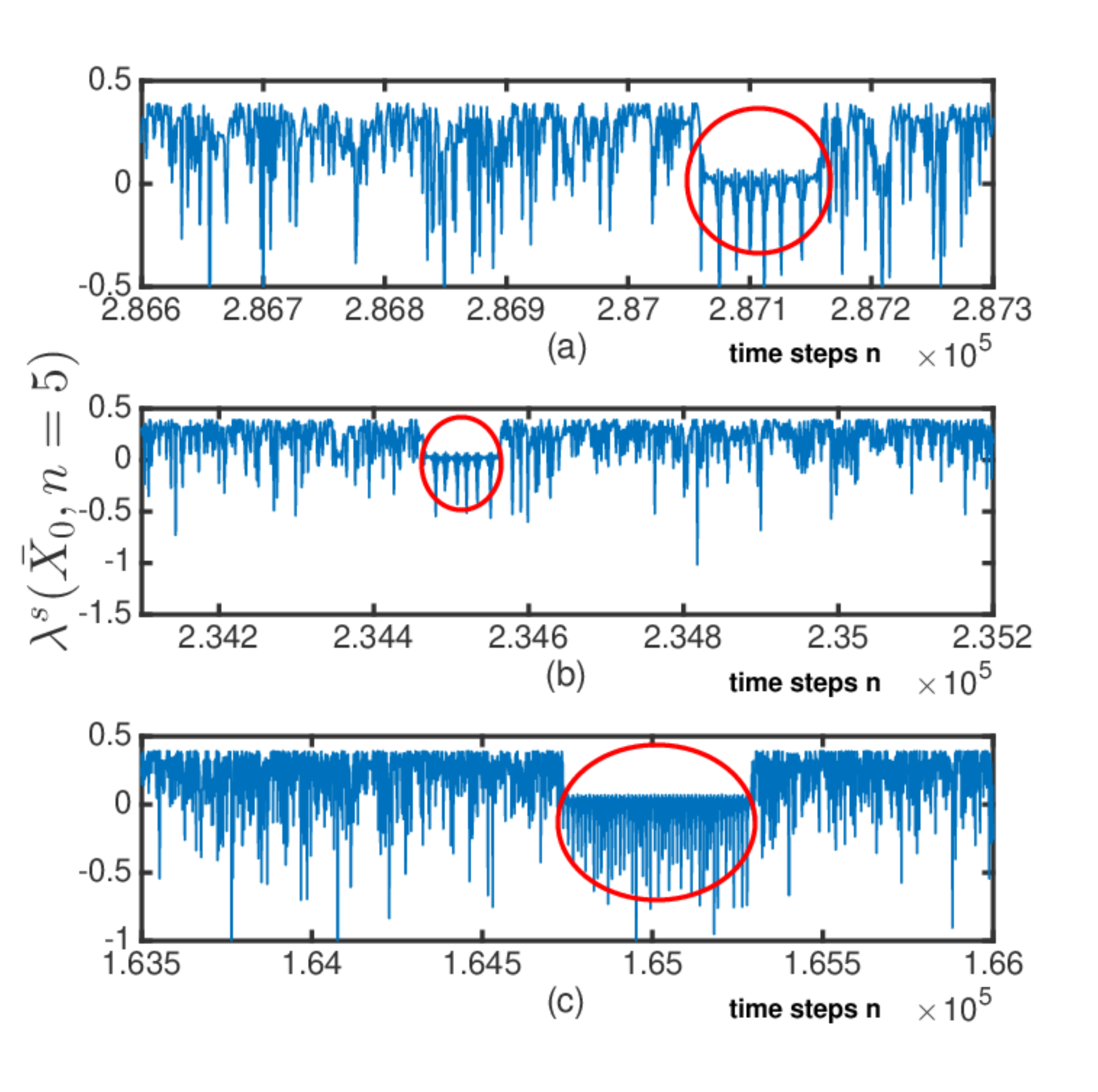}
	\caption{\label{fig:FTLE}\footnotesize Fluctuations of time-5 FTLEs for three different set of initial conditions -- $(P^d_0,Q^d_0,Q^r_0) = (0.466,0.098,0.141)$ in (a),  $(0.009,0.026,0.129)$ in (b), and $(0.553,0.554,0.396)$ in (c). The synchronization times in each cases are 287090, 234521, and 164782 respectively. The time scale on the $x$-axis indicates iteration steps $t$ at which time-5 FTLEs have been computed. The drop window in (a), (b), and (c) shown in ellipses in red occur around synchronization times. Synchronization of chaotic trajectories in the coupled system under study typically occurs in one of such traps.}
\end{figure*}
We now write the variational equation for Eq.(\ref{equ:trans}) by linearizing 
about $(P^\textnormal{d}_n,Q^\textnormal{d}_n)$
\begin{equation}
\label{equ:variational}
\left[ \begin{array}{c} \delta P^\perp_{n+1} \\ \delta Q^\perp_{n+1} 
\end{array} \right] = \mathcal{M}(\alpha)\left[ \begin{array}{c}\delta 
P^\perp_{n} \\ \delta Q^\perp_{n} \end{array} \right],
\end{equation}
where the matrix $\mathcal{M}(\alpha)$ is given by
\begin{equation} 
\label{equ:M}
\mathcal{M}(\alpha) = \begin{bmatrix} 1-\alpha & K\cos(2\pi Q^\textnormal{d}_n) \\ 1 - 
\alpha & 1 + K\cos(2\pi Q^d_n) \end{bmatrix}.
\end{equation}

This is the master stability equation for the unidirectionally coupled 
standard map. The variational equation (\ref{equ:variational}) is the master 
stability equation for the coupled system under investigation. The associated 
largest Lyapunov exponent (LE) computed from the master stability equation is the master stability function (MSF) of the system, given by:
\begin{equation}
\label{equ:msf}
\lambda  = \lim_{n \rightarrow \infty} \lim_{\delta\overbar{X}_0\rightarrow 
0}\frac{1}{n}\sum^{n-1}_{i=0}\ln|JM^n(\overbar{X}_i)|.
\end{equation}

Here $n$ is a positive integer and $JM^n(\overbar{X}_i)$ denotes the Jacobian matrix of the $n$-times iterated map.  A negative value of the MSF (the largest non-zero LE) will ensure that $(P^\perp,Q^\perp) $ tend to zero indicating that the difference between $P$ and $Q$ will die out and the system will synchronize. Now, for the Pecora-Carroll approach, we set $\alpha = 1$. This substitution simplifies the matrix in Eq.(\ref{equ:M}) which now reads

\begin{equation} 
\mathcal{M}(1)= \begin{bmatrix} 0 & K\cos(2\pi Q^\textnormal{d}_n) \\ 0 & 1 + K\cos(2\pi 
Q^d_n) \end{bmatrix},
\end{equation}

The MSF should then be computed from the eigenvalues of the matrix $\mathcal{M}(1)$. It is easy to see that, for $\alpha =1$, one of the eigenvalues of   $\mathcal{M}(1)$ is zero. Therefore, we need to consider only the non-zero eigenvalue which is $1+K\cos(2\pi Q^d_n)$. The corresponding LE is given by

\begin{equation}
\label{equ:LE}
\lambda   = \lim_{n \rightarrow \infty} 
\frac{1}{n}\sum^{n-1}_{i=0}\ln|1+K\cos(2\pi Q^\textnormal{d}_i)|.
\end{equation}

In general, we define the $kth$ time-$n$ LE associated with an initial point $\overbar{X}_0 = (P_0,Q_0)$ for a map $M(P,Q)$ as
\begin{equation}
\label{equ:FLEf}
\lambda _k(\overbar{X}_0;n)  = 
\frac{1}{n}\sum^{n-1}_{i=0}\ln|JM^n(\overbar{X}_i)|.
\end{equation}

Here $n$ is a positive integer and $JM^n(\overbar{X}_i)$ denotes the Jacobian matrix of the $n$-times iterated map. We extend this notion to the master stability function defined in the Sec.~\ref{Master stability function}  i.e.  the non-zero LE defined in Eq.(\ref{equ:LE}) 
which takes the following finite time version 
\begin{equation}
\lambda_1^s(\overbar{X}_0;n) = \frac{1}{n}\sum^{n-1}_{i=0}\ln|1 + K \cos (2\pi 
Q_i)|.
\end{equation}

The subscript ($k = 1$) has been dropped hereafter as we have only one 
exponent to compute.

We plot the variation in the time-5 finite time Lyapunov exponent (FTLE) 
defined above at $K = 6$. At the point of synchronization, the FTLE values 
attain a set of smaller values consistently in a small window, indicating the 
existence of a synchronization trap.  In Fig.~\ref{fig:FTLE} , the three plots 
show the fluctuations  of FTLE near the point of synchronization for three 
different sets of initial conditions - $(P^d_0,Q^d_0,Q^r_0) = 
(0.466,0.098,0.141)$ in (a), $(0.009,0.026,0.129)$ in (b), and 
$(0.553,0.554,0.396)$ in (c) with 
synchronization times 287090, 234521, and 164782 respectively. The values on 
the $x$-axis indicate time steps at which averages are computed so that 
synchronization times and the temporal neighborhood are effectively captured 
in the plots wherein a temporary drop is clearly visible, indicated by 
ellipses in red. Synchronization of chaotic trajectories typically occurs in 
one of such traps. For details for the  synchronization process, see 
~\cite{Mahata2016,Das2017}.

\section{Delayed Synchronization Times}
\label{sec:delay}
\begin{figure}[h]
	\includegraphics[scale=0.45]{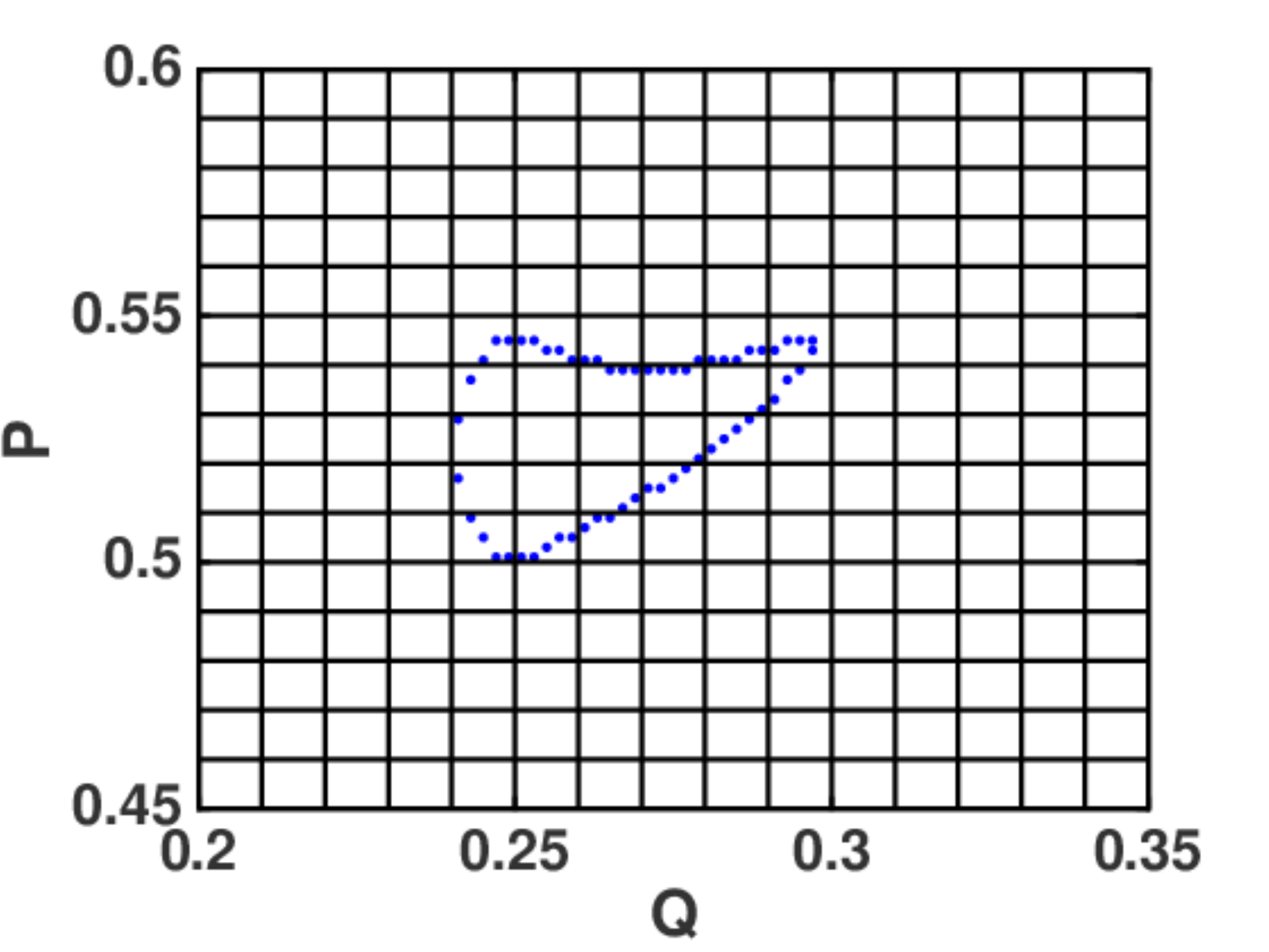}
	\caption{\label{fig:edge_reg_island} \footnotesize Example to detect the 
		edge of a regular island in the phase space using edge detection 
		algorithm  due to Benkadda {\it et al.} \cite{Benkadda1997}. }
\end{figure}

\begin{figure}[b]
	\includegraphics[scale=0.45]{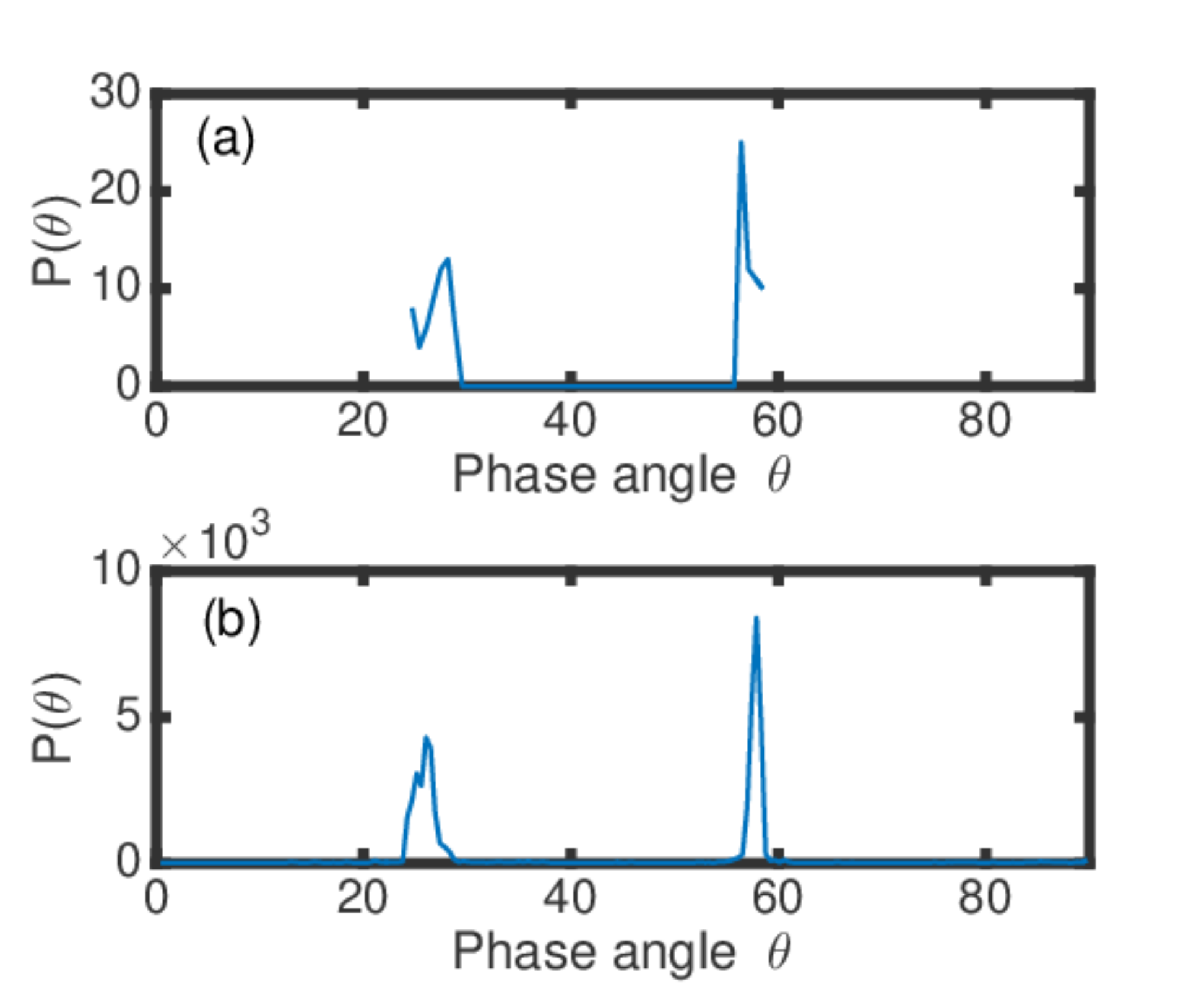}
	\caption{\label{fig:Edge_Sync_angles} \footnotesize Numerical detection of 
		synchronization traps (a) Distribution of phase 
		angles of the points on the edges of regular islands determined by the edge 
		detection algorithm. (b) Distribution of phase angles for points where 
		synchronization occurs. }
\end{figure}

In order to develop a numerical technique to delay synchronization, we  first 
have to identify the vicinity of regular islands in the phase space. A 
mechanism  to suppress stickiness based on the knowledge of hyperbolic and non 
hyperbolic regions in the phase has been reported recently~\cite{Kruger2015}. 
Our method, however, targets the trajectories themselves in the sticky region 
to delay the process.  It is to be noted that the proposed procedure ignores 
the details of a rather complex hierarchy of structures around regular 
islands. We employ the edge-detection algorithm due to Benkadda {\it et al.} 
\cite{Benkadda1997} to identify the edges of regular islands in phase space. 
For the numerical procedure, we divide the phase space in $100 \times 100$ 
grid. The edges of both islands are detected by applying the standard map 
to point initiated in the chaotic sea, for $10^7$ times.  The edges thus 
detected is shown in Fig~\ref{fig:edge_reg_island}.  We have chosen the 
proximity parameter $d=0.02$ i.e. Euclidean distance from points 
on the edge indicate the extent of the vicinity of the island.  This vicinity, 
therefore, indicate the domain wherein the synchronization traps exist.  To 
demonstrate this explicitly, we compare the phase angles, defined by $\theta = 
\tan^{-1}(\frac{Q}{P}$), of the points on numerically detected edges and 
the points of synchronization, as follows.
In Fig.~\ref{fig:Edge_Sync_angles}(a), we show the distribution of phase angles of the points on the edges of regular islands determined by the edge-detection algorithm.  Bimodality of the distribution is due to the existence of two sharp islands in otherwise chaotic bulk in the phase space.  We compare this with the phase angles corresponding to the points in the phase space at which synchronization of the chaotic trajectories occurs.

\begin{figure}[h]
    \includegraphics[scale=0.42]{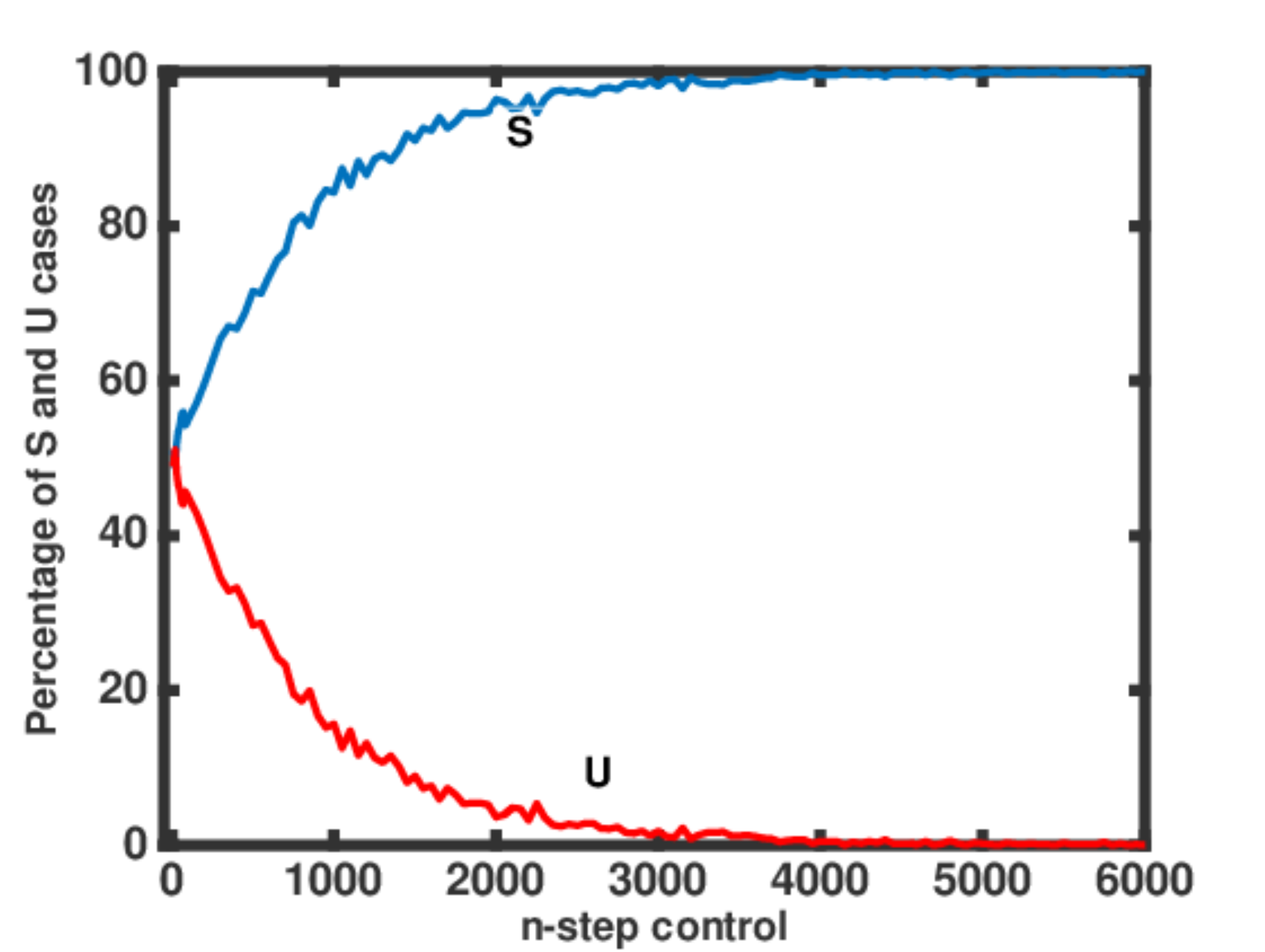}
    \caption{\label{fig:Control_success}\footnotesize The curves indicate the 
        percentages of (S) successful delays shown as the blue curve, and (U) 
        undesirable fast synchronization shown as the red curve. }
\end{figure}

Fig.~\ref{fig:Edge_Sync_angles}(b) shows the distribution of these phase angles for about 50 000 randomly chosen initial conditions for the drive and response maps that lead to synchronization. The locations of the peaks in the distribution shown in ~\ref{fig:Edge_Sync_angles}(a) approximately matches with this in ~\ref{fig:Edge_Sync_angles}(b). It is, therefore, visibly clear that synchronization occurs in the same angular domain of the phase space where the regular islands exist. Our numerically detected edges, thus correctly locate the synchronization traps in the phase space.  We now discuss the control mechanism to delay synchronization time. 

The proximity parameter $d_0 = 0.02$ defines the region in the neighborhood of 
the numerically detected edges wherein synchronization traps exist. The basic 
idea to control synchronization time is to identify the step at which the 
chaotic trajectory visits the numerically determined sticky neighborhood 
followed by a slight deflection so that the trajectory restarts at a point 
outside.  Our procedure depends upon how many times we deflect the trajectory  
which will be referred to as step control. This means that if an $n-$step 
control is  employed then, the trajectory will be kicked away from the domain 
$n$-times during its first $n$ visits i.e. once per visit. The deflection is 
achieved by adding a randomly generated fraction below $0.1$ to the point 
visiting the domain, and therefore the maximum deflection area is roughly 
$1\%$ of the phase space.  This procedure applied on a given set of initial 
conditions of the drive and response maps, may result in four possibilities of 
synchronization time -- (1) successful delay  \textbf{(S)}, (2) no delay 
\textbf{(N)}, (3) failed to achieve synchronization \textbf{(F)}, and (4) 
undesirable faster synchronization \textbf{(U)}. Clearly, the only first 
possibility is the aim of the mechanism; the last one constitute a hazard 
while the other two are trivial. 

For numerical implementation of the mechanism, we have considered $1000$ 
randomly chosen initial conditions from the chaotic sea of the phase space at 
$K = 6$. The maps have been iterated for $10^7$ times in each case.  The 
sticky domain is determined by the proximity parameter $d_0$ around the 
numerically detected edges of the regular islands. We demonstrate the 
effectiveness of the control procedure  with respect to several $n-$step 
controls. The computations have been performed for $n \in \mathbb{A}$, where 
$\mathbb{A} = \{10, 20,30,...,90, 100, 150,200, ... ,10 000\}$.  

\begin{figure}[t]
	\includegraphics[scale=0.42]{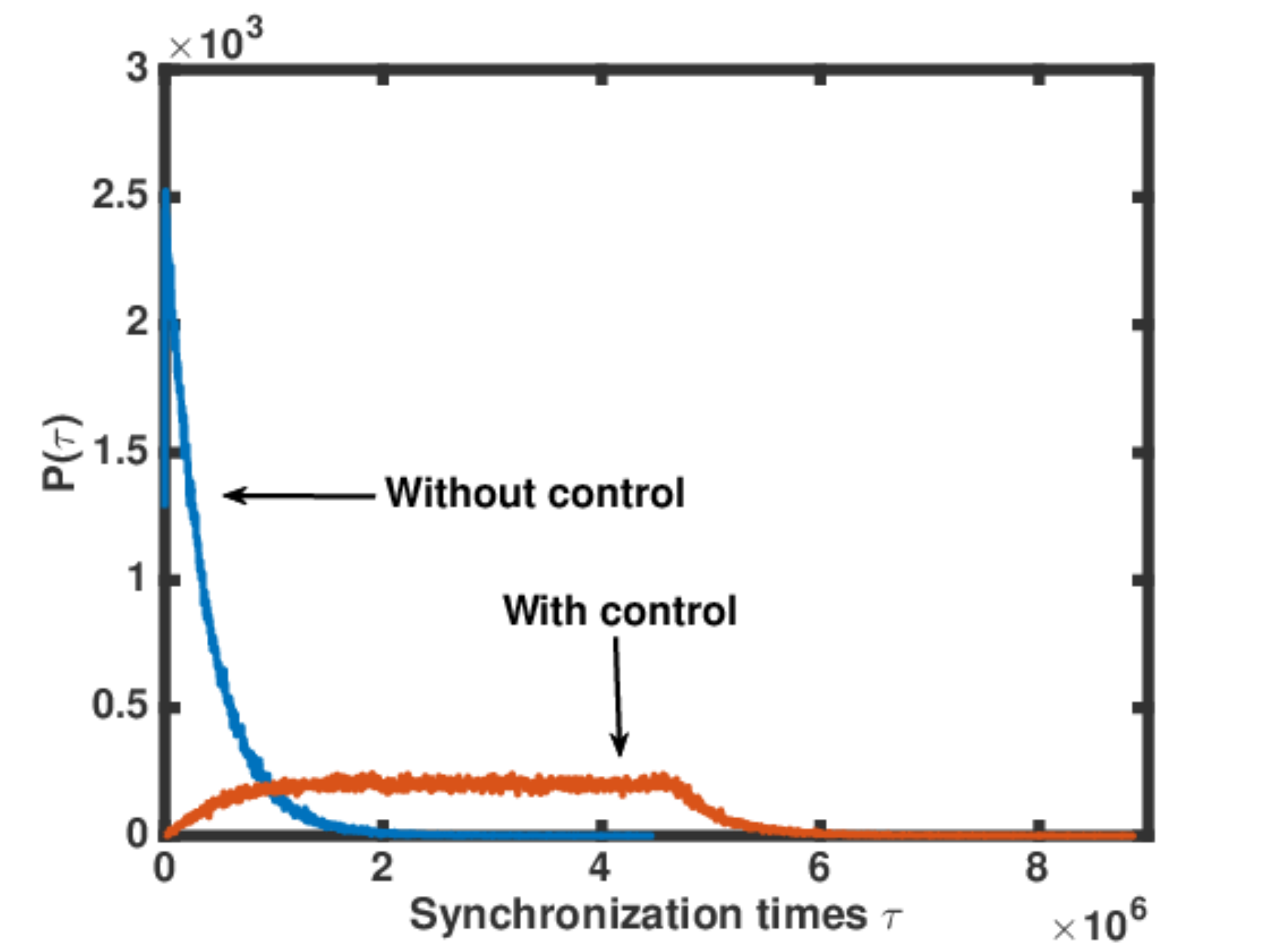}
	\caption{\label{fig:Sync_time_dist}\footnotesize Distribution of 
		synchronization times without n-step control (the blue curve) and with 
		n-step 
		control (the red curve). The control procedure delays synchronization times. }
\end{figure}

\begin{figure}[b]
    \includegraphics[scale=0.42]{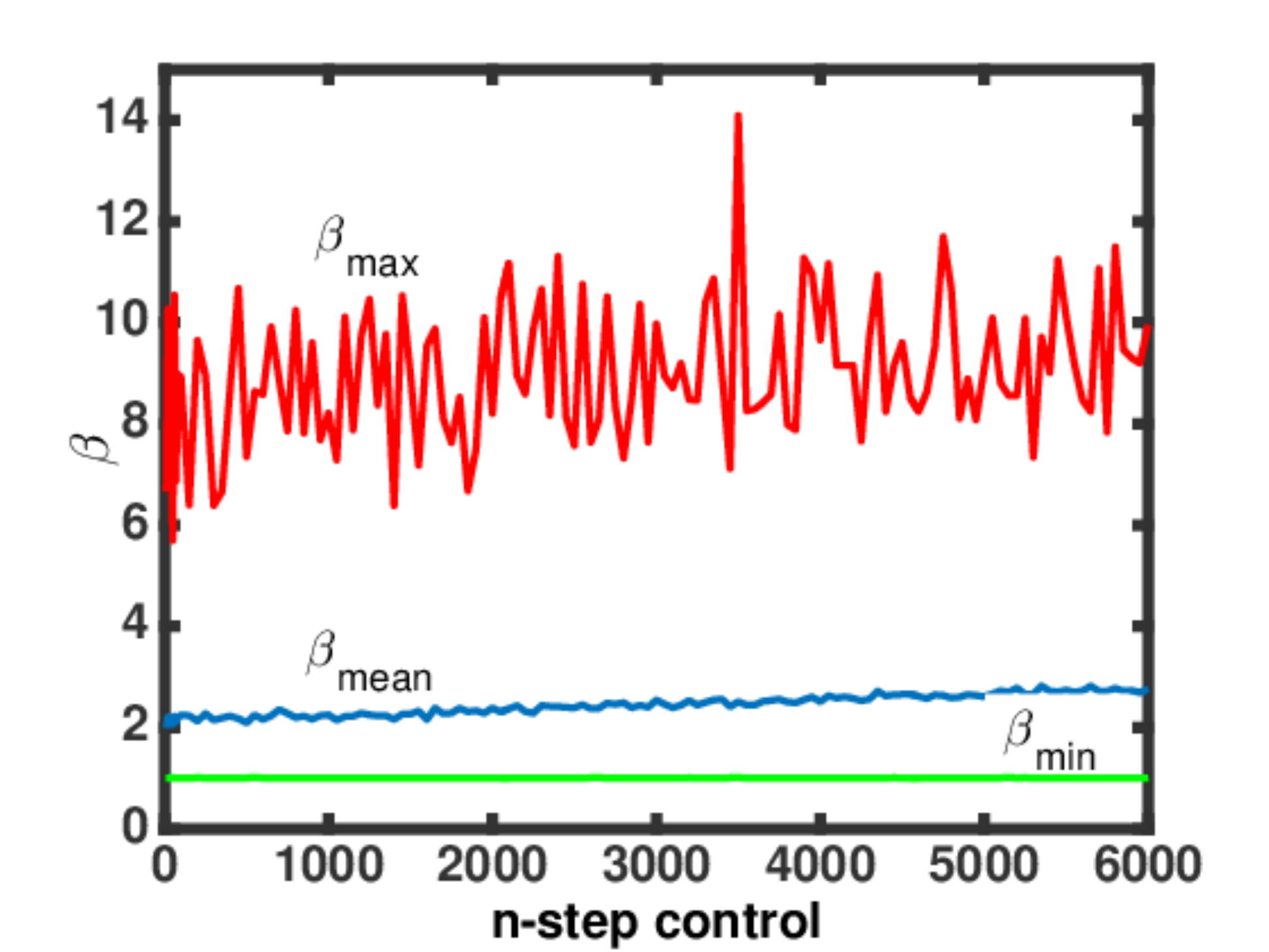}
    \caption{\label{fig:Strength_con}\footnotesize Variation of $\beta$ 
        (maximum, minimum and average) which measures the strength of 
        successful 
        delays i.e. $\beta = \ln(\frac{\tau_d}{\tau_0}$). $\beta_{max}$ 
        indicates the maximum delay time obtained at a give n-step control 
        while $\beta_{mean}$ stands for the mean value around 2. This means 
        that on an average, synchronization times re delayed by 100 times. The 
        minimum value of $\beta$ remains 1 for the cases when no delay was 
        obtained. }
\end{figure}

\begin{figure*}[t]
    \includegraphics[scale=1]{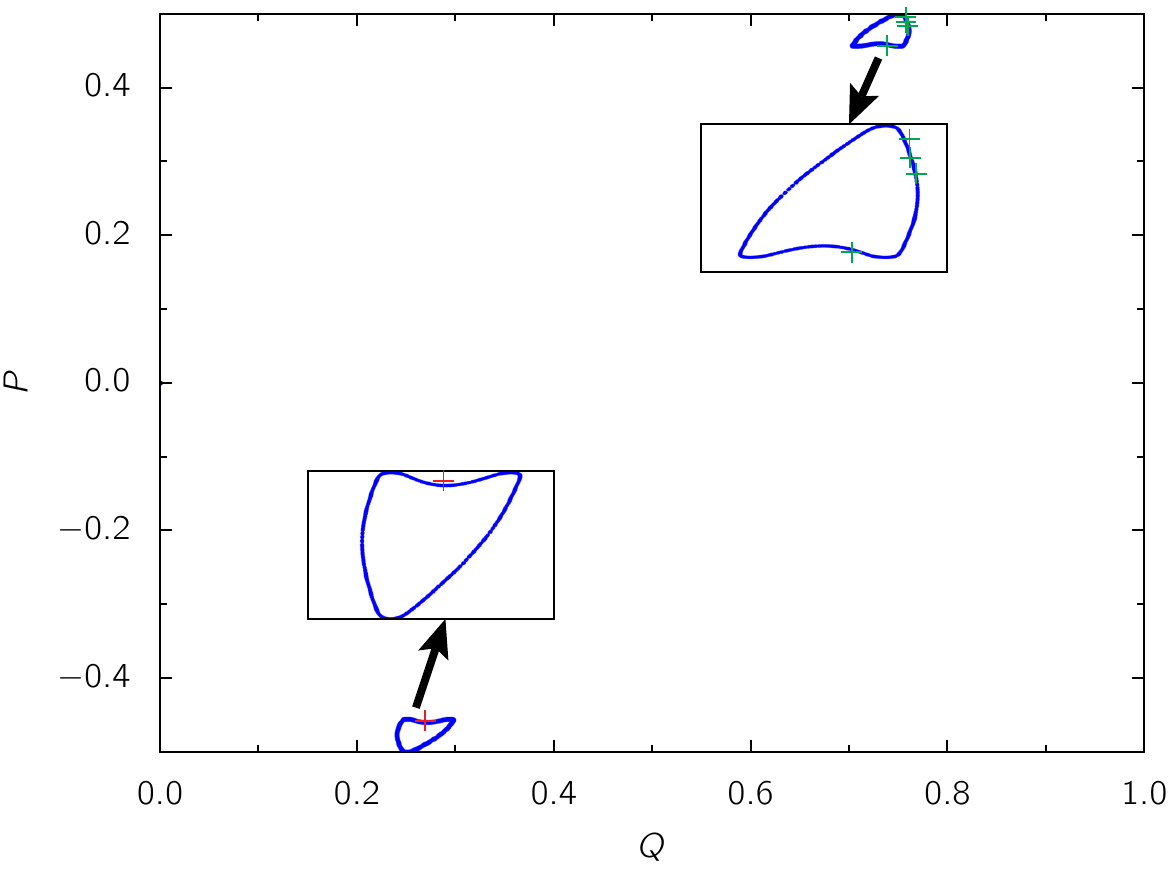}
    \caption{\label{fig:location}\footnotesize The phase space of a sample
    drive map at K = 6 (with $\epsilon = 0.35$) displaying last 1500 
    iterates  until synchronization takes place. The point of synchronization 
    in the neighborhood of the lower island indicated by the sign `+' in color 
    red.  The plot also 
        shows, for illustration, other points of synchronization by the sign 
        `+' in color green for four other sample drive trajectories. The 
        insets are magnified views of the two islands as indicated. Clearly, 
        the 
        parameter perturbation forces trajectories to the neighborhood of the 
        regular islands where synchronization occurs.}
\end{figure*}
Fig.~\ref{fig:Control_success} shows the percentage of successful delays 
\textbf{(S)} (on the y-axis) for a given $n-$step control (on the $x$-axis) up 
to $n = 5500$ after which the successful cases are always $100\%$.  We achieve 
about $50\%$ successful cases for $n = 10$, which rises rapidly to $99\%$ for 
$n = 2950$. It is to be noted that the average synchronization time without 
any control, $\tau_{avg} \sim 10^5$. Therefore, the n-steps required to obtain 
all successful delays is just about $3\%$ of $\tau_{avg}$. We also plot the 
distribution of synchronization times for successful cases and compare it 
against that of corresponding times without control in 
Fig~\ref{fig:Sync_time_dist}. The long-tailed 
distribution of times without delay crosses over to a distribution which shows 
large flat region before a tail. The flat region indicate large number of 
successfully applied control with typical delay times of the order of $10^6$. 
For completeness, we show the decay in the number of undesired cases 
\textbf{U} in Fig~\ref{fig:Control_success}.  The cases \textbf{F} and 
\textbf{N} which are largely present for $n<10$, are extremely small in number 
and are neglected.

The degree of controlled synchronization $\beta$ may be defined as the ratio 
of synchronization times without numerical control, $\tau_c$ to 
synchronization times $\tau_0$ without control, i.e. $\beta =  
\ln(\frac{\tau_d}{\tau_0}$).  Therefore, for successfully delayed 
synchronization $\beta>0$ indicating $\tau_c > \tau$. The plot in 
Fig.~\ref{fig:Strength_con} shows the variation of $\beta$ i.e. $\beta_{min}$, 
$\beta_{mean}$, and $\beta_{max}$  indicating minimum, mean, and maximum 
$\beta$ for considered $n-$step controls. Clearly, we are able to attain 
$\beta_{max}$ values to be in the range $5$ to $14$ while mean values remain 
$\sim2$. Therefore, we are successful in achieving 
significantly large delays in synchronization times.

There are a couple of limitations of the proposed control procedure. The deflection to the trajectory visiting the sticky domain should not insert it inside the regular island. The regular trajectories of the drive and response map are known to synchronize significantly faster and would result in the undesirable possibility \textbf{(U)}. Therefore, an estimate of the size of regular islands, in addition to their locations must be known. These may be determined efficiently, as we have shown, using the edge-detection algorithm. The deflection area  may then be chosen suitably.  Another limitation is that the procedure is unable to impose a pre-determined delay to synchronization times. The random deflections simply keep the trajectories temporarily away from the synchronization traps in the sticky domain and synchronization may eventually occur after the removal of the control.

\section{Advanced Synchronization Times}
\label{sec:advanced}
We now explore the possibility to reduce synchronization times 
i.e. to expedite the process of synchronization. Once again, we will make use 
of the fact that synchronization traps typically exist in the sticky 
neighborhoods of the regular islands.  The procedure is based on a parameter 
perturbation technique used to generate coherent structures in the phase space 
of area-preserving maps~\cite{Gupte2007}. However, the technique may also be 
used to push chaotic trajectory to the neighborhood of periodic islands where 
synchronization traps exit.  We describe the procedure briefly.

We take the standard map as in Eq.~\ref{equ:stdmap} 
with a modulo operation such that $-0.5 \leq P_n \leq 0.5, 0\leq Q_n \leq 1$.
In this form, the standard map has a hyperbolic fixed point at 
$(P_\textnormal{f},Q_\textnormal{f}) = (0,0)$,$(0,0.5)$. We perturb the 
parameter $K$ to $K-\epsilon$ if $|P - P_\textnormal{f}| 
< \delta$, $|Q-Q_\textnormal{f}| < \delta$. For $P$ and $Q$ outside this 
$\delta$-strip, 
$K$ does not change.  It may be noted that the perturbed standard map remains 
to be area-preserving. The Jacobian $J$ is now given by: 

\begin{minipage}[t]{0.45\textwidth}
	\centering
	\[ J = \left( \begin{array}{cc}
	1 + (K-\epsilon)\cos(2\pi Q_n) & 1   \\
	(K-\epsilon)\cos(2\pi Q_n) & 1 \end{array} \right),\]
\end{minipage}

\vspace*{.2cm}

\noindent where, $ \epsilon \neq 0$ if $|P_f - P_n| < \delta$, $|Q_f - Q_n| < 
	\delta$\\ $\epsilon = 0$, otherwise.\\

The determinant of this matrix $J$ remains 1. The phase space thus obtained 
for $K = 6$ , $\delta = 0.4$ and $\epsilon = 0.35$ for the fixed point 
$(P_\textnormal{f},Q_\textnormal{f})=(0.0,0.0)$ is shown in 
Fig.~\ref{fig:location}.   We have also performed computation for the other 
fixed point and with other modulo operations as well and results do not differ 
significantly. Our choice here of modulo operation and the fixed point 
$(P_\textnormal{f},Q_\textnormal{f})$ has some advantage over the 
others in terms of actual numerical values of synchronization times and the 
display of locations of the points where synchronization occurs.

\begin{figure}[t]
	\includegraphics[scale=0.8]{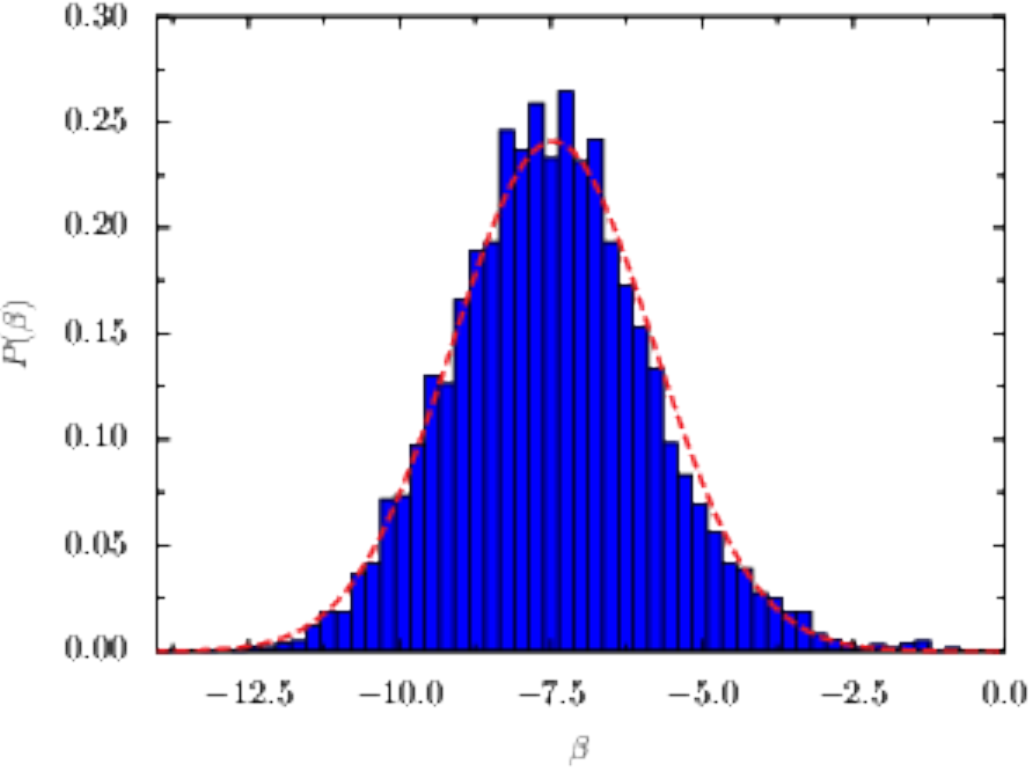}
	\caption{\label{fig:beta_dist}\footnotesize Distribution of degree of 
		controlled synchronization, $P(\beta)$, mean is $-4.69$ and standard 
		deviation $1.72$.}
\end{figure}

For the computations, we take 5 000 initial conditions chosen randomly from 
the chaotic region of the unperturbed standard map in the region $P_0 \in \{-0.4,0.4\}$ 
and $Q_0 \in \{0.3,0.7\}$. The total number of iterations used for each 
initial condition to check for synchronization is $10^7$ and threshold of 
synchronization is $10^{-5}$ as previously.  The procedure has 
been applied to both drive and 
response map in accordance with our coupling  scheme described above. We 
choose a patch around $(P_\textnormal{f},Q_\textnormal{f})$  of length 
$\delta= 0.4$ which corresponds 
to about $16\%$ of the phase space area. We successfully obtain advance 
synchronization in more than $99\%$ of the cases among which a  couple of 
instances  has been shown in Fig.~\ref{fig:location} -- synchronization occurs 
in the neighborhood of one of the period 2  islands. The plots show the 
iterates of a sample drive trajectory until the point of synchronization 
(indicated by `+' in green) is reached. Other `+' signs in red indicate, for 
illustration, points of synchronization in four other cases in order to show 
that it occurs in the sticky neighborhoods.

We again define the degree of controlled synchronization as $\beta = \ln 
(\frac{\tau_c}{\tau}$). For advanced synchronization, $\beta < 0$ indicating 
that synchronization times upon parameter perturbation is smaller than in the
normal course (without any perturbation) i.e. $\tau_c < \tau$. The normalized 
probability distribution of 
$\beta$ in shown in Fig.~\ref{fig:beta_dist}, which may be fit with a Gaussian 
of the following form:
\begin{equation}
f(\beta) =  
\frac{1}{\sqrt{2\pi\sigma^2}}\exp\Big(-\frac{(\beta-\mu)^2}{2\sigma^2}\Big).
\end{equation}

We estimate the mean value $\mu \sim -4.69$ and standard deviation $\sigma 
\sim1.72$ which corresponds to the fact that, on an average, advanced 
synchronization times thus obtained are about $10^2$ times smaller than the 
normal synchronization times with $95.5\%$ values are between $3$ to $3 \times 
10^3$ times smaller.  Therefore, the synchronization times we obtained upon 
parameter perturbation is significantly smaller than the normal 
synchronization times. We would like to point that for a larger 
perturbation  will be more efficient in reducing synchronization times. For 
example, at $\epsilon = 2.0$ synchronization times are largely
from $10$ to $10^5$ times smaller.

Thus, this parameter perturbation technique is an efficient way to reduce
synchronization times with high efficiency.  The value of the perturbation 
imposed should be chosen suitably as a very small may not have any influence 
on the synchronization process. The procedure also depends on the size and 
location of the patch wherein the control is applied.

\section{Discussions and Conclusions}
\label{sec:conclusions}

To summarize, we have developed numerical procedures to control 
synchronization in two identical standard maps coupled under the 
drive-response configuration. The procedure is based on the fact that 
synchronization in the system typically occurs in the sticky neighborhoods of the regular islands in 
the phase space. We have shown that a delay in synchronization can be obtained 
by kicking the trajectories slightly away from the domain temporarily. The 
efficiency of the procedure depends upon the number of steps employed in 
control. By applying the procedure to the system on merely  $1\%$ of the phase 
space area (at $K = 6$), we have successfully increased synchronization times. 
The maximum number of control steps that are used is $0.01\%$ of the total of 
number of iterations used throughout the computation.  We showed that the 
procedure gives variety of delays. The limitations of this procedure are -- 
(a) the deflection may insert the drive trajectory in the regular islands 
which would lead to undesirable faster synchronization, and (b) a 
predetermined synchronization time in not obtainable. 

Furthermore, we have demonstrated a simple way to reduce synchronization 
times in the coupled system. Our technique targets the neighborhood of the 
periodic islands to obtain rapid synchronization. The distribution of degree 
of controlled synchronization times indicates the efficiency of the procedure 
--  synchronization times are reduced upto a factor of $10^3$. We point out 
that the other techniques~\cite{Ott1990, Lai1993b}, in principle, 
could be applied to obtain synchronization if the trajectory are contained in 
synchronization traps long enough for synchronization to take place. In 
both the numerical procedures used in this work, we have ignored the complex 
hierarchical structures which is the origin of the sticky behavior of chaotic 
orbits in the phase 
space.

A possible future direction is to investigate the role of stickiness in 
phenomenon of measure synchronization (MS) in coupled Hamiltonian systems. We 
expect that location of sticky neighborhoods in the phase space may have an 
influence on the transitions involving different partial MS, complete MS, and 
desynchronization states~\cite{Gupta2017}. 
Moreover, we note that the mechanism of parameter perturbation rapidly locates 
small regular islands in the mixed phase with a large chaotic component. 
Locations of these rather small periodic structures in a otherwise chaotic sea 
is thus determined quickly using only a small number of sample trajectories 
within a few hundred iterations (see Fig.~\ref{fig:location}). The classical 
problem of targeting invariant structures in Hamiltonian 
system~\cite{Schroer1997, Cartwright2002} may be revisited if we understand 
the mechanism in detail and adapt it appropriately. As far as experimental 
situations are concerned, the stickiness is a known phenomenon with 
interesting consequences in a variety of systems such as intramolecular energy
redistribution~\cite{Sethi2012}, microwave ionization of Rydberg 
atoms~\cite{Benenti2000} etc. In these contexts, a control
method based on the location of the sticky neighborhoods could be useful.

\begin{acknowledgments}
    \vspace*{-0.3cm}
I am grateful to Neelima Gupte for discussions and Arnd B\"acker for
suggestions to improve the manuscript. I am also thankful to the anonymous 
reviewers for their useful comments.
\end{acknowledgments}

\appendix

\bibliographystyle{cpg_unsrt_title_for_phys_rev}

\bibliography{Syncref}

\begin{thebibliography}{10}
\newcommand{\enquote}[1]{``#1''}
\providecommand{\url}[1]{\texttt{#1}}
\providecommand{\urlprefix}{URL }
\providecommand{\eprint}[2][]{\url{#2}}

\bibitem{Klafter1994}
J.~Klafter and G.~Zumofen, \emph{L\'evy statistics in a {Hamiltonian} system},
  Phys. Rev. E \textbf{49}, 4873 (1994).

\bibitem{Zaburdaev2015}
V.~Zaburdaev, S.~Denisov, and J.~Klafter, \emph{L\'evy walks}, Rev. Mod. Phys.
  \textbf{87}, 483 (2015).

\bibitem{Zaslavsky2002a}
G.~Zaslavsky, \emph{Chaos, fractional kinetics, and anomalous transport}, Phy.
  Rep. \textbf{371}, 461  (2002).

\bibitem{Zaslavsky2002b}
G.~Zaslavsky, \emph{Dynamical traps}, Physica D: Nonlinear Phenomena
  \textbf{168--169}, 292  (2002).

\bibitem{Mackay1984}
R.~Mac{K}ay, J.~Meiss, and I.~Percival, \emph{Transport in {Hamiltonian}
  systems}, Physica D: Nonlinear Phenomena \textbf{13}, 55  (1984).

\bibitem{Easton1993}
R.~W. Easton, J.~D. Meiss, and S.~Carver, \emph{Exit times and transport for
  symplectic twist maps}, Chaos \textbf{3} (1993).

\bibitem{Meiss2015}
J.~D. Meiss, \emph{Thirty years of turnstiles and transport}, Chaos
  \textbf{25}, 097602 (2015).

\bibitem{Altmann2006}
E.~G. Altmann, A.~E. Motter, and H.~Kantz, \emph{Stickiness in {Hamiltonian}
  systems: From sharply divided to hierarchical phase space}, Phys. Rev. E
  \textbf{73}, 026207 (2006).

\bibitem{Altmann2010}
E.~G. Altmann and A.~Endler, \emph{Noise-enhanced trapping in chaotic
  scattering}, Phys. Rev. Lett. \textbf{105}, 244102 (2010).

\bibitem{Livorati2012}
A.~L.~P. Livorati, T.~Kroetz, C.~P. Dettmann, I.~L. Caldas, and E.~D. Leonel,
  \emph{Stickiness in a bouncer model: A slowing mechanism for {Fermi}
  acceleration}, Phys. Rev. E \textbf{86}, 036203 (2012).

\bibitem{Bunimovich2012}
L.~A. Bunimovich and L.~V. Vela-Arevalo, \emph{Many faces of stickiness in
  {Hamiltonian} systems}, Chaos \textbf{22}, 026103 (2012).

\bibitem{Kruger2015}
T.~S. Kr\"uger, P.~P. Galuzio, T.~d.~L. Prado, R.~L. Viana, J.~D. Szezech, and
  S.~R. Lopes, \emph{Mechanism for stickiness suppression during extreme events
  in {Hamiltonian systems}}, Phys. Rev. E \textbf{91}, 062903 (2015).

\bibitem{Lange2016}
S.~Lange, A.~B\"acker, and R.~Ketzmerick, \emph{What is the mechanism of
  power-law distributed {Poincar\'e} recurrences in higher-dimensional
  systems?}, EPL \textbf{116}, 30002 (2016).

\bibitem{Babiano1994}
A.~Babiano, G.~Boffetta, A.~Provenzale, and A.~Vulpiani, \emph{Chaotic
  advection in point vortex models and two‐dimensional turbulence}, Phys.
  Fluids \textbf{6} (1994).

\bibitem{Tel2005}
T.~T{\'e}l, A.~de~Moura, C.~Grebogi, and G.~Károlyi, \emph{Chemical and
  biological activity in open flows: A dynamical system approach}, Phys. Rep.
  \textbf{413}, 91  (2005).

\bibitem{Szezech2012}
J.~D. Szezech, I.~L. Caldas, S.~R. Lopes, P.~J. Morrison, and R.~L. Viana,
  \emph{Effective transport barriers in nontwist systems}, Phys. Rev. E
  \textbf{86}, 036206 (2012).

\bibitem{Martins2014}
C.~G.~L. Martins, M.~Roberto, and I.~L. Caldas, \emph{Delineating the magnetic
  field line escape pattern and stickiness in a poloidally diverted tokamak},
  Phys. Plasmas \textbf{21}, 082506 (2014).

\bibitem{Efthymiopoulos1999}
C.~Efthymiopoulos, G.~Contopoulos, and N.~Voglis, \emph{Cantori, islands and
  asymptotic curves in the stickiness region}, Celest. Mech. Dyn. Astron.
  \textbf{73}, 221 (1999).

\bibitem{Harsoula2010}
M.~Harsoula, C.~Kalapotharakos, and G.~Contopoulos, \emph{Asymptotic orbits in
  barred spiral galaxies}, Mon. Notices Royal Astron. Soc. \textbf{411}, 1111
  (2011).

\bibitem{Harsoula2016}
M.~Harsoula, C.~Efthymiopoulos, and G.~Contopoulos, \emph{Analytical forms of
  chaotic spiral arms}, Mon. Notices Royal Astron. Soc. \textbf{459}, 3419
  (2016).

\bibitem{Mahata2016}
S.~Mahata, S.~Das, and N.~Gupte, \emph{Synchronization in area-preserving maps:
  Effects of mixed phase space and coherent structures}, Phys. Rev. E
  \textbf{93}, 062212 (2016).

\bibitem{Szezech2005}
J.~{Szezech Jr.}, S.~Lopes, and R.~Viana, \emph{Finite-time lyapunov spectrum
  for chaotic orbits of non-integrable {Hamiltonian} systems}, Phy. Lett. A
  \textbf{335}, 394  (2005).

\bibitem{Pecora1998}
L.~M. Pecora and T.~L. Carroll, \emph{Master stability functions for
  synchronized coupled systems}, Phys. Rev. Lett. \textbf{80}, 2109 (1998).

\bibitem{Boccaletti2002}
S.~Boccaletti, J.~Kurths, G.~Osipov, D.~Valladares, and C.~Zhou, \emph{The
  synchronization of chaotic systems}, Phys. Rep. \textbf{366}, 1 (2002).

\bibitem{Pecora2015}
L.~M. Pecora and T.~L. Carroll, \emph{Synchronization of chaotic systems},
  Chaos \textbf{25}, 097611 (2015).

\bibitem{Hampton1999}
A.~Hampton and D.~H. Zanette, \emph{Measure synchronization in {Coupled}
  {Hamiltonian} systems}, Phys. Rev. Lett. \textbf{83}, 2179 (1999).

\bibitem{Wang2003}
X.~Wang, Z.~Ying, and G.~Hu, \emph{Controlling {Hamiltonian} systems by using
  measure synchronization}, Phys. Lett. A \textbf{298}, 383 (2002).

\bibitem{Vincent2005}
U.~E. Vincent, \emph{Measure synchronization in coupled {Duffing Hamiltonian}
  systems}, New J. Phys. \textbf{7}, 209 (2005).

\bibitem{Gupta2017}
S.~Gupta, S.~De, M.~S. Janaki, and A.~N. Sekar~Iyengar, \emph{Exploring the
  route to measure synchronization in non-linearly coupled {Hamiltonian}
  systems}, Chaos \textbf{27}, 113103 (2017).

\bibitem{Das2017}
S.~Das, S.~Mahata, and N.~Gupte, \emph{Synchronization, phase slips, and
  coherent structures in area-preserving maps}, Proceedings of the Conference
  on Perspectives in Nonlinear Dynamics - 2016 \textbf{1}, 205 (2017).

\bibitem{Grabow2011}
C.~Grabow, S.~Grosskinsky, and M.~Timme, \emph{Speed of complex network
  synchronization}, Eur. Phys. J. B \textbf{84}, 613 (2011).

\bibitem{Wang2016}
S.-J. Wang, R.-H. Du, T.~Jin, X.-S. Wu, and S.-X. Qu, \emph{Synchronous slowing
  down in coupled logistic maps via random network topology}, Sci. Rep.
  \textbf{6}, 1 (2016).

\bibitem{Thorpe1996}
S.~Thorpe, D.~Fize, and C.~Marlot, \emph{Speed of processing in the human
  visual system}, Nature \textbf{381}, 520 (1996).

\bibitem{Uchida2003}
N.~Uchida and Z.~F. Mainen, \emph{Speed and accuracy of olfactory
  discrimination in the rat}, Nature Neuroscience \textbf{6}, 1224 (2003).

\bibitem{Boccaletti2000}
S.~Boccaletti, C.~Grebogi, Y.-C. Lai, H.~Mancini, and D.~Maza, \emph{The
  control of chaos: theory and applications}, Phys. Rep. \textbf{329}, 103
  (2000).

\bibitem{Ott2008}
E.~Ott, \emph{Controlling chaos}, Scholarpedia \textbf{1}, 1699 (2008).

\bibitem{Bolotin2009}
Y.~Bolotin, A.~Tur, and V.~Yanovsky, \emph{Chaos: Concepts, Control and
  Constructive Use}, Understanding Complex Systems, Springer Berlin Heidelberg
  (2009).

\bibitem{Ott1990}
E.~Ott, C.~Grebogi, and J.~A. Yorke, \emph{Controlling chaos}, Phys. Rev. Lett.
  \textbf{64}, 1196 (1990).

\bibitem{Lai1993a}
Y.-C. Lai and C.~Grebogi, \emph{Synchronization of chaotic trajectories using
  control}, Phys. Rev. E \textbf{47}, 2357 (1993).

\bibitem{Lai1994}
Y.-C. Lai and C.~Grebogi, \emph{Synchronization of spatiotemporal chaotic
  systems by feedback control}, Phys. Rev. E \textbf{50}, 1894 (1994).

\bibitem{Nagai1996}
Y.~Nagai, X.-D. Hua, and Y.-C. Lai, \emph{Controlling on-off intermittent
  dynamics}, Phys. Rev. E \textbf{54}, 1190 (1996).

\bibitem{Lai1993b}
Y.-C. Lai, M.~Ding, and C.~Grebogi, \emph{Controlling {H}amiltonian chaos},
  Phys. Rev. E \textbf{47}, 86 (1993).

\bibitem{Chirikov1960}
B.~V. Chirikov, \emph{Resonance processes in magnetic traps}, The Soviet
  Journal of Atomic Energy \textbf{6}, 464 (1960).

\bibitem{Izraelev1980}
F.~Izraelev, \emph{Nearly linear mappings and their applications}, Physica D:
  Nonlinear Phenomena \textbf{1}, 243  (1980).

\bibitem{Chirikov1989}
B.~V. Chirikov and V.~V. Vecheslavov, \emph{{Chaotic dynamics of comet
  {Halley}}}, Astron. Astrophys. \textbf{221}, 146  (1989).

\bibitem{Chirikov2008}
B.~Chirikov and D.~Shepelyansky, \emph{{C}hirikov standard map}, Scholarpedia
  \textbf{3}, 3550 (2008).

\bibitem{Pecora1990}
L.~M. Pecora and T.~L. Carroll, \emph{Synchronization in chaotic systems},
  Phys. Rev. Lett. \textbf{64}, 821 (1990).

\bibitem{Benkadda1997}
S.~Benkadda, S.~Kassibrakis, R.~B. White, and G.~M. Zaslavsky,
  \emph{Self-similarity and transport in the standard map}, Phys. Rev. E
  \textbf{55}, 4909 (1997).

\bibitem{Gupte2007}
N.~Gupte and A.~Sharma, \emph{Creation of coherent structures in
  area-preserving maps}, Phys. Lett. A \textbf{365}, 295 (2007).

\bibitem{Schroer1997}
C.~G. Schroer and E.~Ott, \emph{Targeting in {H}amiltonian systems that have
  mixed regular/chaotic phase spaces}, Chaos: An Interdisciplinary Journal of
  Nonlinear Science \textbf{7}, 512 (1997).

\bibitem{Cartwright2002}
J.~H.~E. Cartwright, M.~O. Magnasco, and O.~Piro, \emph{Bailout embeddings,
  targeting of invariant tori, and the control of {H}amiltonian chaos}, Phys.
  Rev. E \textbf{65}, 045203 (2002).

\bibitem{Sethi2012}
A.~Sethi and S.~Keshavamurthy, \emph{Driven coupled morse oscillators:
  visualizing the phase space and characterizing the transport}, Mol. Phys.
  \textbf{110}, 717 (2012).

\bibitem{Benenti2000}
G.~Benenti, G.~Casati, G.~Maspero, and D.~L. Shepelyansky, \emph{Quantum
  {P}oincar\'e {R}ecurrences for a hydrogen atom in a microwave field}, Phys.
  Rev. Lett. \textbf{84}, 4088 (2000).

\end{thebibliography}

\end{document}